\documentclass[reqno,12pt]{article}
\usepackage{amsmath,amsfonts,amssymb,amsthm,amstext,amscd,eucal}
\usepackage[all]{xy}
\newcommand{\mathsym}[1]{{}} 
\usepackage[colorlinks=true,
            linkcolor=blue,
            urlcolor=blue,
            citecolor=blue]{hyperref}
\usepackage{enumerate}
\usepackage{slashed}
\DeclareMathAlphabet{\pazocal}{OMS}{zplm}{m}{n}
\usepackage{graphicx}% Include figure files
\usepackage{bm}% bold

\usepackage{hyperref}

\usepackage{graphicx}

\makeatletter \@addtoreset{equation}{section}

\makeatletter\renewcommand\section{\@startsection {section}{1}{\z@}%
                                   {-3.5ex \@plus -1ex \@minus -.2ex}%nn
                                   {2.3ex \@plus.2ex}%
                                   {\normalfont\large\bfseries}}
\renewcommand\subsection{\@startsection{subsection}{2}{\z@}%
                                     {-3.25ex\@plus -1ex \@minus -.2ex}%
                                     {1.5ex \@plus .2ex}%
                                     {\normalfont\bfseries}}

\DeclareMathAlphabet{\pazocal}{OMS}{zplm}{m}{n}

\DeclareMathAlphabet{\mathcal}{OMS}{cmsy}{b}{n}

\makeatother

\newcommand{\email}[1]{\footnote{E-mail: \href{mailto:#1}{#1}}}

\begin{document}

\title{\bf\Large{ On the gauge invariance of the higher-derivative Yang-Mills-Chern-Simons action}}

\author{\textbf{M.~Ghasemkhani\email{m$_{_{-}}$ghasemkhani@sbu.ac.ir } $^{a}$, G.~Soleimani\email{g.soleimani.m76@gmail.com} $^{a}$
 and R.~Bufalo\email{rodrigo.bufalo@ufla.br} $^{b}$}\\\\
%EndAName
\textit{\small$^{a}$Department of Physics, Shahid Beheshti University, 1983969411, Tehran, Iran}\\
\textit{\small$^{b}$Departamento de F\'isica, Universidade Federal de Lavras,}\\
\textit{\small Caixa Postal 3037, 37200-900 Lavras, MG, Brazil}\\
}

\maketitle
\date{}

\begin{abstract}

In this work, we study the perturbative generation of the gauge invariant effective action for the non-Abelian gauge field in a $(2+1)$-dimensional spacetime.
We present a detailed analysis of the two, three and four-point functions in order to determine the non-Abelian Chern-Simons terms (parity odd) and Yang-Mills terms (parity even).
Moreover, these terms are supplemented by the higher-derivative corrections which resulted in the Alekseev-Arbuzov-Baikov effective action (parity even) plus the higher-derivative (HD) corrections to the Chern-Simons terms (parity odd).
In order to highlight some features about the perturbative generation of the effective action, we present a discussion based on the dimensional analysis, which allows us to establish the general structure of the permissible terms to guarantee the gauge invariance of the higher-derivative parts.

\end{abstract}

%\vspace{0.5in}

%%%%%%%%%%%%%%%%%%%%%%%%%%%%%%%%%%%%%%%%%%%%%%%%%%%%%%%%%%%%%%%%%%%%%%%%%%%%%%%%%%%%%%%%%%
\setcounter{footnote}{0}
\renewcommand{\baselinestretch}{1.05}  %Line spacing
%%%%%%%%%%%%%%%%%%%%%%%%%%%%%%%%%%%%%%%%%%%%%%%%%%%%%%%%%%%%%%%%%%%%%%%%%%%%%%%%

%\addtocontents{toc}{\protect\setcounter{tocdepth}{2}}
\newpage
%\tableofcontents
%%%%%%%%%%%%%%%%%%%%%%%%%%%%%%%%%%%%%%%%%%%%%%%%%%%%%%%%%%%%%%%%%%%%%%%%%%%%%%%%%%%%%%%%%%%

\section{Introduction}

Undoubtedly, lower dimensional field theory models always have received considerable attention due to the plethora of rich phenomena they describe as well as serving as laboratories where we can learn useful things about the well-recognized four-dimensional problems, ranging from solvable models in two dimensions, to planar physics and condensed matter in three dimensions.

One can observe that the area of lower dimensional field theory models had its beginning with the proposals of the  Thirring \cite{Thirring:1958in} and Schwinger \cite{Schwinger:1962tp} models as exact solvable models in $(1+1)$ dimensions. This class of models also displays the important feature of bosonization \cite{Stone:1995ys}, which has also been applied in different contexts \cite{Yao:2020dqx,Gabai:2022vri,CorsoBSantos:2023ewk}.
One can also note the interest in two-dimensional gravitational solvable models of black holes \cite{Mertens:2022irh}.

On the other hand, in condensed matter physics \cite{Marino:2017ckg,Jackiw:1983nv,Semenoff:1984dq}, three-dimensional models had its dawn in the 1980's in the description of planar phenomena, especially, mass generation in topological models \cite{Schonfeld:1980kb,Deser:1981wh,Redlich:1983kn,Deser:1999pa}, quantum Hall effect \cite{Dunne:1998qy,Susskind:2001fb,Hansson:2016zlh,Iengo:1991zbc}, topological insulators \cite{Hasan:2010xy}, among others.
Moreover, in planar physics, mainly due to the presence of parity violation which is responsible to generate a gauge invariant mass term for the photon, interesting features are present in the photon's dynamics.

A powerful tool and important element in uncovering modifications to quantum field dynamics is the effective action \cite{Buchbinder:1992rb,Dittrich:2000zu,Buchbinder:2021wzv,Quevillon:2018mfl,Bonora:2016otz,Bonora:2017ykb}.
For the case of the (Abelian and non-Abelian) gauge field a good amount of attention was the subject of several studies, in which the dynamical generation of the effective action by the radiative correction was considered at zero temperature ($T=0$) framework \cite{Deser:1999pa,Dunne:1998qy,Bufalo:2014ooa,Ghasemkhani:2019jjy,Arvanitakis:2015oga} as well as at finite temperature case \cite{Deser:1997nv,Deser:1997gp,Fosco:1997ei,Fosco:1997vu}.
Going beyond the leading order, we obtain the higher-derivative corrections to the Chern-Simons term, which are known to modify the physical propagating modes.
These contributions appear in the perturbative expansion of the effective action in terms of
($\frac{p^{2}}{m^2}$) in the large fermion mass limit ($m\gg p$), which were considered in the full Abelian action \cite{Deser:1999pa,Lisboa-Santos:2023pwc}, also in the noncommutative framework \cite{Bufalo:2014ooa}.

Interestingly, the presence of higher derivative terms change the physical behavior of three dimensional field theories.
In the case of topologically massive gravity, the additional (ghost) degree of freedom is eaten by the conformal invariance \cite{Deser:1981wh}.
On the other hand, in the case of (higher derivative) extended Chern-Simons theory, it is no longer sensitive to ``large'' gauge aspects present in the CS model, it is no longer topological but depends explicitly on the background geometry \cite{Deser:1999pa}.

Although many aspects of the three-dimensional Yang-Mills Chern-Simons theory are well established \cite{Pisarski:1985yj,Burgess:1989id,Li:2021vve}, we wish to explore the issue of dynamical generation of higher-derivative terms and couplings to the Yang-Mills Chern-Simons theory.
We follow the effective action approach because all of the generated terms are well-defined and gauge invariant.
In the parity even sector, it is expected to obtain the known Alekseev-Arbuzov-Baikov terms \cite{Alekseev:1981fu,Quevillon:2018mfl}, while in the parity odd sector we should get the usual Deser-Jackiw term \cite{Deser:1999pa} plus corrections to the (novel) cubic self-coupling.

Hence, our goal is to examine the perturbative generation of higher-derivative terms in the three-dimensional Yang-Mills Chern-Simons theory. Moreover, in order to clarify some aspects about the effective action, we consider two types of matter fields, fermionic and scalar fields.
We start our study by considering the interaction between the gauge and fermionic fields in Sec.~\ref{sec2}, in which we evaluate the higher-derivative terms up to the 4-point function contributions to the effective action, which have parity odd and parity even sectors.
Furthermore, in Sec.~\ref{sec3}, we consider the effective action arising from the coupling between the gauge and scalar fields,  which does not present parity odd terms due to the coupled scalar field. Also,we shall see that the parity-even terms have different coefficients than the fermionic counterpart.
In Sec.~\ref{sec4}, we discuss, by means of complementarity, the general structure of higher-derivative terms allowed by gauge invariance using a dimensional analysis procedure, both parity-even and parity-odd sectors are examined.
We present our final remarks and perspectives in Sec.~\ref{sec5}.

%%%%%%%%%%%%%%%%%%%%%%%%%%%%%%%%%%%%%
\section{Fermionic matter coupled to the non-Abelian gauge field}
\label{sec2}
%%%%%%%%%%%%%%%%%%%%%%%%%%%%%%%%%%%%%

In this section, we consider the fermionic matter minimally coupled to an external non-Abelian gauge field. The relevant action is described by the following
\begin{equation}
S_{F}=\int d^{3}x~\bar\Psi\left(i\slashed{D}-m\right)\Psi,
\label{eq:2-1}
\end{equation}
where $D_{\mu}=\partial_{\mu}-igG_{\mu}^{a}T^{a}$ is the covariant derivative. Also, $G_\mu^a$ are the gauge fields and $T^{a}$ are the $SU(N)$ generators.
The one-loop effective action $\Gamma_{\rm eff}$ for the gauge
field is readily obtained by integrating out the fermionic fields as below
\begin{equation}
e^{i\Gamma_{\rm eff}[G]}=\int D\bar\psi D\psi~e^{iS_{F}}.
\label{eq:2-2}
\end{equation}
Applying the usual grassmann Gaussian functional integration formulas, the 1PI effective action can formally be written as a perturbative series \footnote{It should be realized the presence of a normalization factor here in order to satisfy  $\Gamma_{\rm eff}[0]=1$, but we omit it for the simplicity of notation.}
\begin{equation}
\Gamma_{\rm eff}[G]=-i{\rm Tr} \ln\left(i\slashed{D}-m\right)=-\sum_{a=1}^{N^2-1}\sum_{n=1}^{\infty}\frac{g^{n}}{n} {\rm Tr}\Big[\Big(\frac{1}{i\slashed{\partial}-m}\Big)\slashed{G}^{a} T^{a}\Big]^{n},
\label{eq:2-3}
\end{equation}
here, the trace ${\rm Tr}$ is a sum over both Dirac ($Tr$) and color (${\rm tr}$) indices \footnote{ For the first time, Dyson realized that the number of diagrams of $n$-th order typically increases as $n!$, which suggested that the perturbation series is divergent \cite{Dyson}. A useful mathematical technique for handling the divergent series is the Borel summation to improve the convergence of the series. Moreover, as we know, interchanging the order of a double sum is permissible when the double summation converges absolutely, but, here is not the case. Nevertheless, we use a suitable regularization method to allow us for interchanging the double summation \cite{Weinberg, Hardy, Dunne-2008}.}

 Also, the effective action \eqref{eq:2-3} is equivalent to the following form
\begin{equation}
\Gamma_{\rm eff}[G]=\sum_{n=1}^{\infty}\int d^{3}x_{1}\ldots\int d^{3}x_{n}~G_{a_{1}}^{\mu_{1}}(x_{1})\ldots\ G_{a_{n}}^{\mu_{n}}(x_{n})~\Pi_{\mu_{1}\ldots\mu_{n}}^{a_{1}\ldots a_{n}}\left(x_{1},\ldots,x_{n}\right),
\label{eq:2-4}
\end{equation}
in which $\Pi_{\mu_{1}\ldots\mu_{n}}^{a_{1}\ldots a_{n}}\left(x_{1},\ldots,x_{n}\right)$ represents the $n$-point function of the non-Abelian gauge field with the minimal coupling to the fermionic matter.
From a diagrammatic point of view, it includes the one-loop graphs contributing to the gauge field $n$-point functions which are considered as
\begin{equation}
\Gamma_{\rm eff}[G]=\mathcal{S}_{\rm eff}[G]+\mathcal{S}_{\rm {eff}}[GG]+\mathcal{S}_{\rm eff}[GGG]+\mathcal{S}_{\rm eff}[GGGG]+\cdots.
\end{equation}
In order to examine the terms of interest in the effective action (the Chern-Simons and Yang-Mills terms, as well as its higher-derivative corrections), we shall now proceed in evaluating explicitly the one-loop contributions of \eqref{eq:2-4} for two,
three and four-point function of the gauge field. Here, we remark that the first term of the series \eqref{eq:2-4}, corresponding to the one-point function (tadpole contribution), vanishes due to the traceless property of the $SU(N)$ generators i.e. $\mathcal{S}_{\rm eff}[G]=0$. After this, we set $N=3$ in our detailed analysis which yields the one-loop effective action for the gluons.

%%%%%%%%%%%%%%%%%%%%%%%%%%%%%%%%%%%%%%%%%%%%%%%%
\subsection{GG-term contribution}
%%%%%%%%%%%%%%%%%%%%%%%%%%%%%%%%%%%%%%%%%%%%%%%%

\begin{figure}[t]
\vspace{-1.2cm}
\includegraphics[height=2.7\baselineskip]{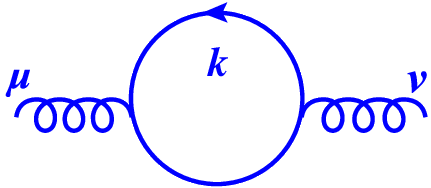}
 \centering\caption{Fermionic one-loop graph contributing to $GG$-term.}
\label{fig:oneloop1}
\end{figure}

Let us start with the $n=2$ contribution of the series \eqref{eq:2-4} which corresponds to the free part of the gluon effective action.
The Feynman amplitude of the one-loop graph, depicted in Fig.~\ref{fig:oneloop1}, is explicitly given by the following expression
\begin{equation}
\Pi_{\mu\nu}^{ab}(p)=-g^2\int\frac{d^{3}k}{\left(2\pi\right)^3}\frac{Tr\left[\gamma_{\mu}(\slashed{k}+\slashed{p}+m)
\gamma_{\nu}(\slashed{k}+m)\right]}{(k^2-m^2)((k+p)^2-m^2)}~{\rm tr}\left(T^{a}T^{b}\right).
\end{equation}
With help of the trace identities in $1+2$ dimensions \cite{Deser:1981wh}
\begin{equation}
Tr(\gamma^{\mu}\gamma^{\nu})=2\eta^{\mu\nu}, \quad Tr(\gamma^{\mu}\gamma^{\nu}\gamma^{\rho})=2i\epsilon^{\mu\nu\rho},\quad  Tr(\gamma^{\mu}\gamma^{\nu}\gamma^{\rho}\gamma^{\sigma})=2(\eta^{\mu\nu}\eta^{\rho\sigma}-\eta^{\mu\rho}\eta^{\nu\sigma}+\eta^{\mu\sigma}\eta^{\nu\rho}),
\end{equation}
and ${\rm tr}(T^{a}T^{b})=\frac{1}{2}\delta^{ab}$, as well as dimensional regularization, in the low-energy limit, $p^2\ll m^2$, we have
\begin{equation} \label{eff_GG1}
\Pi_{\mu\nu}^{ab}(p)=g^2 \delta^{ab} \left[\frac{i}{24\pi m}\left(p_\mu p_\nu-p^2\eta_{\mu\nu} \right)-\frac{1}{8\pi}\epsilon_{\mu\nu\rho}p^\rho+\mathcal{O}(m^{-2})\right],
\end{equation}
which clearly satisfies the Ward identity $p^{\mu}\Pi_{\mu\nu}^{ab}(p)=0$. It should be noted that the first term (parity even, orders $m^{-1}$, $m^{-3}$,...) leads to the kinetic term of the Yang-Mills action, while the second term (parity odd, orders $m^{0}$, $m^{-2}$,...) leads to the kinetic term of the non-Abelian Chern-Simons action.

Moreover, we can cast the expression \eqref{eff_GG1} in terms of the effective action
\begin{align} \label{eff_GG}
\mathcal{S}_{\rm{eff}}^{\rm{YM}}[GG]&=-\frac{ig^2}{48\pi m}\int d^3x\left[G_{\mu}^a(x)\left(-\square \eta^{\mu\nu}+\partial^{\mu}\partial^{\nu}\right)G_{\nu}^{a}(x)\right],\\
\mathcal{S}_{\rm {eff}}^{\rm{CS}}[GG]&=\frac{ig^2}{16\pi} \int d^3x \left[\epsilon^{\mu\nu\rho}G_{\mu}^a(x)\partial_{\nu}G_{\rho}^a(x)\right].\label{8}
\end{align}

In the next to leading order, which includes the higher-derivative corrections to the GG-term, we find the HD contributions to the kinetic parts of YM and CS terms as below
\begin{align}
\Pi_{\mu\nu}^{ab}(p)\Big|_{\tiny\mbox{HD-CS}}&=\frac{g^2\delta^{ab}}{96\pi m^2}~\epsilon_{\mu\rho\nu}p^{\rho}p^2+\mathcal{O}(m^{-4}),\label{eq_hd_CS}\\
\Pi_{\mu\nu}^{ab}(p)\Big|_{\tiny\mbox{HD-YM}}&=-\frac{ig^2\delta^{ab}}{240\pi m^3}\left(p^2\eta_{\mu\nu}-p_\mu p_\nu \right)p^{2}+\mathcal{O}(m^{-5}). \label{eq_hd_YM}
\end{align}
Here, we observe the presence of an additional $p^2$ (or $\square$) factor in the effective action.
This is an expected structure, because it preserves the symmetries of the theory.

%%%%%%%%%%%%%%%%%%%%%%%%%%%%%%%%%%%%%%%%%%%
\subsection{GGG-term contribution}
%%%%%%%%%%%%%%%%%%%%%%%%%%%%%%%%%%%%%%%%%%%

\begin{figure}[t]
\vspace{-1.2cm}
\includegraphics[height=6.5\baselineskip]{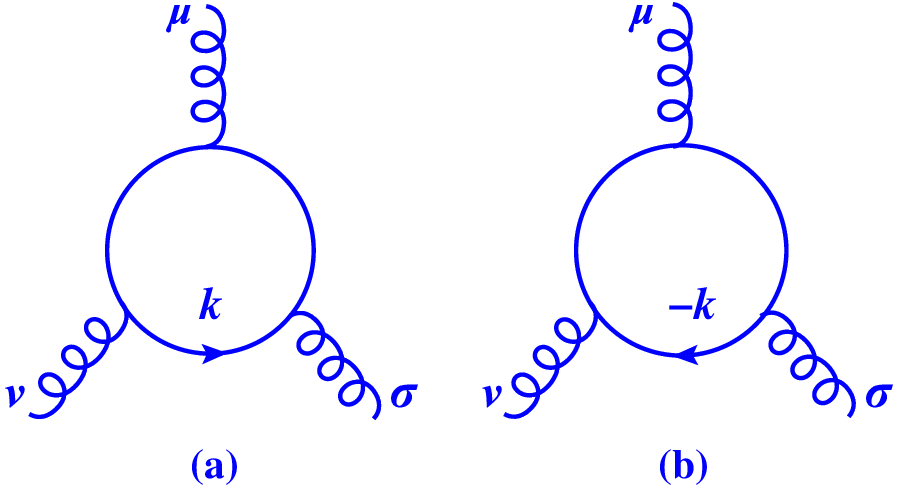}
 \centering\caption{Fermionic one-loop graphs contributing to $GGG$-term.}
\label{fig:oneloop2}
\end{figure}

Based on the perturbative expansion \eqref{eq:2-4}, there are two diagrams contributing to the three-point function, these are depicted in Fig.~\eqref{fig:oneloop2}. The relevant amplitude is given by
\begin{align}
\Pi^{abc}_{\mu\nu\sigma}\Big|_{(a+b)}&=g^3\int\frac{d^{3}k}{\left(2\pi\right)^3}\frac{Tr\left[\gamma_{\mu} \left(\slashed{k}-\slashed{s}+m\right)\gamma_{\nu} \left(\slashed{k}+m\right)\gamma_{\sigma}\left(\slashed{k}+\slashed{r}+m\right)\right]}{(k^2-m^2)((k+r)^2-m^2)((k-s)^2-m^2)}\notag\\
&\times\left[{\rm tr}\left(T^{a}T^{b}T^{c}\right)-{\rm tr}\left(T^{a}T^{c}T^{b}\right)\right].
\label{eq:3-point}
\end{align}
According to the properties of SU(3) generators
\begin{equation}
{\rm tr}\left(T^{a}T^{b}T^{c}\right)=\frac{1}{4}(d^{abc}+if^{abc}),
\end{equation}
in which $f^{abc}$ and $d^{abc}$ are the anti-symmetric and symmetric parts of the structure constants, respectively.

Using the trace of Dirac matrices properties and dimensional regularization to compute the momentum integrals, we obtain
\begin{align}\label{7}
\Pi_{\mu\nu\sigma}^{abc}\Big|_{(a+b)}&=-\frac{g^3f^{abc}}{24 \pi m}\left[\eta_{\mu\sigma}\left(r-p\right)_{\nu}+\eta_{\mu\nu}\left(p-s\right)_{\sigma}+\eta_{\sigma\nu}\left(s-r\right)_{\mu}\right]
+\frac{ig^3}{8\pi}f^{abc}\epsilon_{\mu\nu\sigma}+\mathcal{O}(m^{-2}).
\end{align}
The first part  (parity even) of Eq.\eqref{7} is exactly the vertex coefficient of three gluons in the Yang-Mills theory and the second part  (parity odd) is related to the non-Abelian Chern-Simons action, which are proportional to $m^{-1}$ and $m^{0}$, respectively.

By considering \eqref{8} and the parity odd part of \eqref{7}, we observe that both contributions are of order $m^{0}$. Thus, by combining them, we have the functional action for the ordinary non-Abelian Chern-Simons theory
\begin{equation}  \label{eq_CS2}
\Gamma_{\rm{eff}}^{\rm{CS}}\Big|_{\mathcal{O}(m^{0})}\propto \frac{g^2}{m^{0}}
\int d^3x~\epsilon^{\mu\nu\sigma}\Big[G_{\mu}^{a}(x)\partial_\nu G_\sigma^{a}(x)+\frac{ig}{3}f^{abc}G_\mu^{a}(x)G_\nu^{b}(x)G_\sigma^{c}(x)\Big],
\end{equation}
or equivalently
\begin{equation}\label{eq_CS2-equivalent}
\Gamma_{\rm{eff}}^{\rm{CS}}\Big|_{\mathcal{O}(m^{0})}\propto \frac{2g^2}{m^{0}}
\int d^3x~\epsilon^{\mu\nu\sigma}{\rm tr}\Big[G_{\mu}(x)\partial_\nu G_\sigma(x)+\frac{2g}{3}G_\mu(x)G_\nu(x)G_\sigma(x)\Big],
\end{equation}
which is invariant under the infinitesimal gauge transformations, $\delta G_{\mu}=\partial_{\mu}\lambda-i[G_{\mu},\lambda]$.

It is remarkable that in the Abelian case the amplitude of the graphs contributing to the photon's 3-point function in \eqref{eq:3-point} cancel each other, and hence, the charge conjugation (C) symmetry of QED is also preserved at one-loop level, the Furry's theorem. In QED, this theorem is satisfied at any order through the following identity
 \begin{equation}
\langle \Omega\big|T[j_{\mu_{1}}(x_{1})\cdots j_{\mu_{2n+1}}(x_{2n+1})]\big|\Omega\rangle=0.
\label{eq:furry-abelian}
\end{equation}
The proof of this identity is easily shown by inserting the charge conjugation operator C as ${\rm C}^{\dagger}{\rm C}=1$, as well as using the fact that
 ${\rm C}j_{\mu}{\rm C}^{\dagger}=-j_{\mu}$ and the (vacuum state) invariance ${\rm C}\big|\Omega\rangle=\big|\Omega\rangle$.
On the other hand, in the non-Abelian case, as we observe in \eqref{7}, the relevant amplitude for the vacuum expectation value of 3 currents  does not vanish.
In fact, the main reason for this non-vanishing amplitude refers to the properties of the symmetric group generators.
 To illuminate this point, we remember the behaviour of the $SU(3)$ current, $J_{\mu}^{a}=\bar\psi\gamma_{\mu}T^{a}\psi$ under the charge conjugation (C) in a usual $SU(3)$ gauge theory \cite{ref8}, as below
\begin{equation} \label{eq101}
\big(J_{\mu}^{a}\big)^{\rm C}=\bar\psi(x)\gamma_{\mu}(-T^{a})^{t}\psi(x),
\end{equation}
here, $(T^{a})^{t}$ is the transpose of $T^{a}$. As we observe, the generators $T^a$ are replaced by $(-T^{a})^{t}$ in the conjugate representation which satisfy the following commutation relation
\begin{equation}
\big[(-T^{a})^{t},(-T^{b})^{t}\big]=if^{abc}(-T^{c})^{t},
\end{equation}
which is a closed algebra. The generators $T^{1},T^{3},T^{4},T^{6}$ and $T^{8}$ are symmetric, whereas $T^{2},T^{5}$ and $T^{7}$ are antisymmetric \cite{ref8}.
Thus, the relation \eqref{eq101} can be revised as $\big(J_{\mu}^{a}\big)^{\rm C}=-\xi(a)J_{\mu}^{a}(x)$, in which
\begin{equation}
\xi(a)=\left\{
  \begin{array}{ll}
    +1, & a=1, 3, 4, 6, 8; \\
    -1, & a=2, 5, 7.
  \end{array}
\right.
\end{equation}
Therefore, in order to make a C-invariant interaction term $S_{\rm int}=g\int d^{3}xJ_{\mu}^{a}(x)G^{\mu a}(x)$, it is a sufficient condition to consider that $(G_{\mu})^{\rm C}=-G_{\mu}^{a}(x)(T^{a})^{t}$ or $\big(G_{\mu}^{a}\big)^{\rm C}=-\xi(a)G_{\mu}^{a}(x)$.

Now, if we consider the anti-symmetric generators, i.e. $\xi(a)=\xi(b)=\xi(c)=-1$, we arrive at
\begin{equation}
\langle \Omega\big|T\big[J_{\mu}^{a}(x)J_{\nu}^{b}(y)J_{\rho}^{c}(z)\big]\big|\Omega\rangle \neq 0,
\label{eq:furry}
\end{equation}
which is in contrast to the abelian case.
However, in the case of symmetric generators, i.e. $\xi(a)=\xi(b)=\xi(c)=1$, the amplitude \eqref{eq:furry} vanishes.
As a result, we can conclude that Furry's theorem is not fully satisfied in a non-abelian gauge theory.

Moreover, it is interesting to observe that the induced non-Abelian Chern-Simons action \eqref{eq_CS2-equivalent} under charge-conjugation transformation $(G_{\mu})^{\rm C}=-G_{\mu}^{a}(x)(T^{a})^{t}$ behaves as below
\begin{equation}
\Gamma_{\rm{eff}}^{\rm{CS}}\Big|_{\mathcal{O}(m^{0})}^{\rm C}\propto \frac{2g^2}{m^{0}}\int d^3x~\epsilon^{\mu\nu\sigma}\Big[G_{\mu}^{a}(x)\partial_\nu G_\sigma^{b}(x)~{\rm tr}\big(T^{b}T^{a}\big)^{t}-\frac{2g}{3}G_\mu^{a}(x)G_\nu^{b}(x)G_\sigma^{c}(x)~{\rm tr}\big(T^{c}T^{b}T^{a}\big)^{t}\Big].
\end{equation}
Using ${\rm tr}(\emph{\textbf{A}})={\rm tr}(\emph{\textbf{A}}^{t})$ and the trace identities of SU(3) generators as well as the antisymmetric property of $\epsilon^{\mu\nu\sigma}$, it is easy to show that $(\Gamma_{\rm{eff}}^{\rm{CS}})^{\rm C}=\Gamma_{\rm{eff}}^{\rm{CS}}$ and hence the non-Abelian Chern-Simons action is C-invariant. In order to complete our discussion, we remind that under parity transformation
\begin{equation}
 G_{0}^{a}(t,\vec{x})\longrightarrow G_{0}^{a}(t,-\vec{x}),~~~~G_{i}^{a}(t,\vec{x})\longrightarrow -G_{i}^{a}(t,-\vec{x}),
\end{equation}
 the CS action is parity odd $(\Gamma_{\rm{eff}}^{\rm{CS}})^{\rm P}=-\Gamma_{\rm{eff}}^{\rm{CS}}$. Eventually, we conclude that the non-Abelian Chern-Simons action is CP-odd.

%%%%%%%%%%%%%%%%%%%55

In the next to leading order, the HD contributions to the the parity even part is obtained as
 \begin{align}\label{eq_hd_YM2}
\Pi_{\mu\nu\sigma}^{abc}\Big|_{\tiny\mbox{HD-YM}}=-\frac{g^3 f^{abc}}{1280\pi m^3}&~\bigg\{\eta_{\nu\sigma}\Big[r^2(3r-s)+s^2(r-3s)+8(r.s)(r-s)\Big]_{\mu}\notag\\
&-\eta_{\mu\sigma}\Big[r^2(7r+6s)+s^2(9r+8s)+6(r.s)(2r+3s)\Big]_{\nu}\notag\\
&+\eta_{\mu\nu}\Big[r^2(13r+9s)+s^2(11r-2s)+4(r.s)(7r+3s)\Big]_{\sigma}
+\cdots\bigg\}+\mathcal{O}(m^{-5}),
\end{align}
while to the parity odd it reads
\begin{align} \label{eq_hd_CS2}
\Pi_{\mu\nu\sigma}^{abc}\Big|_{\tiny\mbox{HD-CS}}=\frac{ig^3f^{abc}}{192\pi m^2}&\Big[p^{2}\epsilon_{\mu\nu\sigma}-2p_{\mu} p^{\beta}\epsilon_{\nu\sigma\beta}-4p^{\alpha}s^{\beta}\eta_{\nu\sigma} \epsilon_{\mu\alpha\beta}-2s^{\alpha} s_{\nu}\epsilon_{\mu\alpha\sigma}+2p_{\mu}s^{\beta}\epsilon_{\nu\beta\sigma}\notag\\
&+2p^{\alpha}s_{\nu}\epsilon_{\mu\alpha\sigma}+4(p.s)\epsilon_{\mu\nu\sigma}-2p^{\alpha}s^{\beta}\eta_{\mu\nu}\epsilon_{\beta\sigma\alpha}+\cdots\Big]+\mathcal{O}(m^{-4}),
\end{align}
which are proportional to $m^{-3}$ and $m^{-2}$, respectively.
Here, $"\cdots"$ indicates terms which are other tensorial combinations contributing to both parts in the next to leading order, but are omitted for simplicity of notation.
We shall examine in sec.~\ref{sec4} the (gauge invariant) general structure of such terms by means of dimensional analysis.

%The three-gluon interaction part of Chern-Simons action is obtained as follows
%\begin{align}
%\Gamma_{eff}^{CS}[GGG]=\frac{i\,g^3}{24\pi}\int d^3x f^{abc} \epsilon^{\mu\nu\sigma}G_{\mu}^{a}(x)G_{\nu}^{b}(x)G_{\sigma}^{c}(x).
%\end{align}

%%%%%%%%%%%%%%%%%%%%%%%%%%%%%%%%%%%%%%%%%%%%%%
\subsection{GGGG-term contribution}
%%%%%%%%%%%%%%%%%%%%%%%%%%%%%%%%%%%%%%%%%%%%%%

\begin{figure}[t]
\vspace{-1.2cm}
\includegraphics[height=5.9\baselineskip]{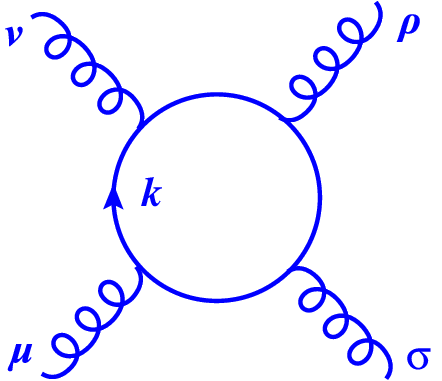}
 \centering\caption{Fermionic one-loop graph contributing to $GGGG$-term.}
\label{fig:oneloop3}
\end{figure}

The S-matrix expansion for the 4-point function includes 24 permutations of the external gluon legs.
Hence, we first compute the amplitude of the one of these graphs (given by fig.~\ref{fig:oneloop3}) and then perform the corresponding permutations.
Hence, the total amplitude is formally written as
\begin{align}
\Pi_{\mu\nu\rho\sigma}^{abcd}=\sum_{i=1}^{24}\Pi_{\mu\nu\rho\sigma}^{abcd}\Big|_{(i)},
\end{align}
and the general diagram can be cast as
\begin{align}
\Pi_{\mu\nu\rho\sigma}^{abcd}\Big|_{(1)}&=-g^4\int\frac{d^3k}{(2\pi)^3}\frac{Tr\left[\gamma_\mu \left(\slashed{k}+m\right)\gamma_\nu \left(\slashed{k}+\slashed{s}+m\right)\gamma_{\rho} \left(\slashed{k}+\slashed{r}+\slashed{s}+m\right)\gamma_\sigma\left(\slashed{k}-\slashed{p}+m\right)\right]}{\left[k^{2}-m^2\right]\left[(k+s)^{2}-m^2\right]\left[(k+s+r)^{2}-m^2\right]\left[(k-p)^{2}-m^2\right]}
\notag\\&\times\Big[\frac{1}{12}\delta^{ab}\delta^{cd}+\frac{1}{8}d^{abe}d^{cde}+\frac{i}{8}d^{abe}f^{cde}+\frac{i}{8}f^{abe}d^{cde}-\frac{1}{8}f^{abe}f^{cde}\Big],
\end{align}
in which we have used the following identity \cite{ref10}
\begin{equation} \label{ident_T}
{\rm tr}\left(T^{a}T^{b}T^{c}T^{d}\right)=\frac{1}{12}\delta^{ab}\delta^{cd}+\frac{1}{8}d^{abe}d^{cde}+\frac{i}{8}d^{abe}f^{cde}+\frac{i}{8}f^{abe}d^{cde}-\frac{1}{8}f^{abe}f^{cde}.
\end{equation}
Moreover, solving the momentum integration and considering the low-energy limit, we obtain
\begin{align} \label{eq_GGGG}
\Pi_{\mu\nu\rho\sigma}^{abcd}\Big|_{(1)}&=-\frac{ig^4}{96 \pi m}\Big[\frac{2}{3}\delta^{ab}\delta^{cd}+d^{abe}d^{cde}+id^{abe}f^{cde}+if^{abe}d^{cde}-f^{abe}f^{cde}\Big] \notag\\
&\times\Big(\eta_{\mu\nu}\eta_{\rho\sigma}-2\eta_{\mu\rho}\eta_{\nu\sigma}+\eta_{\mu\sigma}\eta_{\nu\rho}\Big)+\mathcal{O}(m^{-2}).
\end{align}
Unlike the low-energy amplitudes of GG and GGG terms, there are no parity odd contributions (order $m^{0}$) in \eqref{eq_GGGG}.
Actually, the leading term of the GGGG term in \eqref{eq_GGGG} is at order of $m^{-1}$ (parity even) which is a Yang-Mills term.
Nonetheless, although this amplitude does not produce any Chern-Simons term at order of $m^{0}$, it contributes to the higher-derivative extension of the Chern-Simons action at order $m^{-2}$.

Hence, performing the 24 permutations in the result \eqref{eq_GGGG}, we arrive at the expression
\begin{align} \label{eff_GGGG}
\Pi_{\mu\nu\rho\sigma}^{abcd}=-\frac{ig^4}{8 \pi m}\left[f^{bce} f^{dae}\left(\eta_{\mu\rho}\eta_{\nu\sigma}-\eta_{\mu\nu}\eta_{\rho\sigma}\right)+ f^{bae} f^{dce}\left(\eta_{\mu\rho}~\eta_{\nu\sigma}-\eta_{\mu\sigma}\eta_{\nu\rho}\right)
 + f^{cae}f^{dbe}\left(\eta_{\mu\nu}\eta_{\rho\sigma}-\eta_{\mu\sigma}\eta_{\nu\rho}\right)\right].
\end{align}
This result is precisely the vertex coefficient of the four-gluon interaction term in the Yang-Mills theory, which is consistent with our expectation.

Finally, collecting the $1/m$ terms (parity even) of \eqref{eff_GG},  \eqref{7} and  \eqref{eff_GGGG}, we obtain the Yang-Mills action
\begin{equation} \label{eff_YM}
\Gamma_{\rm eff}^{YM}\Big|_{\mathcal{O}(m^{-1})}\propto \frac{g^2}{m}\int d^{3}x~{\rm tr} \left[F_{\mu\nu}F^{\mu\nu}\right].
\end{equation}
in which $F_{\mu\nu} = \partial_\mu A_\nu - \partial_\nu A_\mu  +i g[A_\mu,A_\nu]$ is the usual field strength tensor.
Besides, it is worth to mention that in three dimensions the ratio $\frac{g^{2}}{m}$ is a dimensionless quantity. Here an important comment is in order. The mass dimension of the coupling constant in three dimensions is $[g]=\frac{1}{2}$ so we can define a dimensionless coupling as $\lambda=\frac{g^{2}}{\Lambda}$, in which $\Lambda$ is the energy scale. Now, if we take the IR limit, i.e. $\Lambda\rightarrow 0$, the coupling $\lambda$ grows largely and hence the theory in IR is strongly coupled, but UV finite.

It is important to emphasize that the effective action of Yang- Mills \eqref{eff_YM} is  Lorentz invariant and SU(3) gauge invariant under the transformation $U=e^{i\theta^{a} T^{a}}$, where the field strength tensor has the following transformation $F_{\mu\nu}\rightarrow UF_{\mu\nu}U^{-1}$.

Furthermore, based on the  above discussion, it is readily obtained that the induced YM action \eqref{eff_YM} is invariant under charge conjugation and parity transformations and hence Yang-Mills action, unlike the CS action, is CP-even.

One interesting remark about the GGGG term is that in the Abelian case, one can show that the low-energy 4-photon amplitude in three dimensions is given by:
\begin{equation}
\Pi^{\mu\nu\rho\sigma}\Big|_{\tiny\mbox{(1)}}^{\tiny\mbox{QED}}=-\frac{i g^4}{6\pi m}\left(\eta^{\mu\nu}\eta^{\rho\sigma}-2\eta^{\mu\rho}\eta^{\nu\sigma}+\eta^{\mu\sigma}\eta^{\nu\rho}\right).
\end{equation}
So that, applying the 24 permutations, we conclude that
\begin{equation}
\Pi^{\mu\nu\rho\sigma}\Big|_{\tiny\mbox{QED}}=\sum_{i=1}^{24}\Pi_{\mu\nu\rho\sigma}^{abcd}\Big|_{\tiny{(i)}}=0.
\end{equation}
as we expected, no self-interaction term is generated to the gauge field in the Abelian regime \cite{Dittrich:2000zu}.

Moreover, in the next to leading order, the HD contributions to the YM and CS parts for the GGGG diagram given in Fig.~\ref{fig:oneloop3} are found as
\begin{align}
\Pi_{\mu\nu\rho\sigma}^{abcd}\Big|_{(a,1)}^{\tiny\mbox{HD-YM}}=-\frac{ig^4}{256\pi m^3}&\left\{\left(\eta_{\mu\nu} \eta_{\rho\sigma}+\eta_{\mu\sigma} \eta_{\nu\rho}\right)\Bigl(-\frac{2}{5}p^2-\frac{16}{5}s^2-\frac{2}{5}r^2+\frac{4}{5}\left(r.s\right)+\frac{4}{5}\left(p.s\right)+\frac{11}{5}
\left(p.r\right)\Bigl)\right.\notag\\&\left. +
\eta_{\mu\rho} \eta_{\nu\sigma}\Bigl(\frac{2}{3}p^2+\frac{16}{3}s^2+\frac{2}{3}r^2-\frac{4}{3}\left(r.s\right)-\frac{4}{3}\left(p.s\right)
-4\left(p.r\right)\Bigl)+\cdots\right\}\notag\\&\times
\bigg(\frac{2}{3} \delta^{ab} \delta^{cd}+ d^{abe} d^{cde}+i d^{abe} f^{cde}+i f^{abe} d^{cde}- f^{abe} f^{cde}\bigg)
+\mathcal{O}(m^{-5}) \label{eq_hd_YM3}
\end{align}
\begin{align} \label{eq_hd_CS3}
\Pi_{\mu\nu\rho\sigma}^{abcd}\Big|_{(a,1)}^{\tiny\mbox{HD-CS}}=\frac{3g^4}{128\pi m^2}
&\left\{-\frac{1}{12}s_{\mu}\epsilon_{\nu\rho\sigma}+\frac{1}{6}r_{\sigma}\epsilon_{\mu\nu\rho}+\frac{1}{6}p_{\rho}\epsilon_{\mu\nu\sigma}-\frac{1}{12} s^{\alpha}\eta_{\nu\rho} \epsilon_{\mu\alpha\sigma}-\frac{1}{6}p^{\alpha}\eta_{\mu\nu}\epsilon_{\rho\sigma\alpha}+\cdots\right\}
\notag\\&\times
\bigg(\frac{2}{3} \delta^{ab} \delta^{cd}+ d^{abe} d^{cde}+id^{abe} f^{cde}+i f^{abe} d^{cde}-f^{abe} f^{cde}\bigg)
+\mathcal{O}(m^{-4}),
\end{align}
which are proportional to $m^{-3}$ and $m^{-2}$, respectively. Here, again $"\cdots"$ indicates that there are many other tensorial combinations contributing to both parts, YM and CS, in the next to leading order.
As one can easily realize, the next to leading order result for the GGGG term has a very complicated form so we omit the final result obtained after performing the 24 permutations.

Actually, the higher-derivative parity even terms \eqref{eq_hd_YM}, \eqref{eq_hd_YM2}  and \eqref{eq_hd_YM3} are part of the Alekseev-Arbuzov-Baikov (AAB) effective action \cite{Alekseev:1981fu}
\begin{equation} \label{AAB}
\mathcal{L}_{\rm AAB}  \propto {\rm tr} \left(D_\mu F^{\mu\nu} D^\sigma F_{\sigma\nu}\right) + {\rm tr} \left(D_\lambda F^{\mu\nu} D^\lambda F_{\mu\nu}\right) - g~ {\rm tr} \left( F_{\mu\nu} F^{\nu\lambda}F_{\lambda\alpha} \eta^{\alpha\mu}\right).
\end{equation}
One can observe that to generate the remaining terms of the AAB action  \eqref{AAB} it is necessary to consider contributions up to $n=6$ in the series \eqref{eq:2-4} to engender the GGGGGG terms.
%%%%%%%%%%%%%%%%%%%%%%%%%%%%%%%%%%%%%%%%%%%%%%%%%%%%%%%%%%%%%%%%%%%%%%%%%
%%%%%%%%%%%%%%%%%%%%%%%%%%%%%%%%%%%%%%%%%%%%%%%%%%%%%%%%%%%%%%%%%%%%%%%%
\section{Scalar matter coupled to the non-Abelian gauge field}
\label{sec3}
%%%%%%%%%%%%%%%%%%%%%%%%%%%%%%%%%%%%%%%%%%%%%%%%%%%%%%%%%%%%%%%%%%%%%%%%
%%%%%%%%%%%%%%%%%%%%%%%%%%%%%%%%%%%%%%%%%%%%%%%%%%%%%%%%%%%%%%%%%%%%%%%%%

In order to highlight some interesting aspects about the process of generating the effective action for the gauge field, we consider now the case of a non-Abelian gauge field coupled to the scalar fields (which possesses further couplings).
This case can be thought as the effective description of the interaction, for instance, among  gluons and squarks.
The effective action follows from the path integral of the charged scalar fields
\begin{equation}
e^{i\Gamma_{\rm eff}[G]}=\int D\Phi^\dag D\Phi~e^{-i\int d^{3}x \, \Phi^\dag\left(\lvert D\rvert^2+m^{2}\right)\Phi},
\end{equation}
which, at one-loop order, results in the gluon's effective action by integrating out the scalar fields
\begin{equation}
i\Gamma_{\rm eff}[G]={\rm tr} \ln \left[-\left(\partial_{\mu}+igG_{\mu}^{a} T^{a}\right)\left(\partial^{\mu}-igG^{\mu b} T^{b}\right)+m^2\right].
\end{equation}
Moreover, the above effective action can be cast into the following perturbative form
\begin{equation}
\Gamma_{\rm eff}[G]=\sum_{n=1}^{\infty}\int d^{3}x_{1}\ldots\int d^{3}x_{n}~G_{a_{1}}^{\mu_{1}}(x_{1})\ldots\ G_{a_{n}}^{\mu_{n}}(x_{n})~\Omega_{\mu_{1}\ldots\mu_{n}}^{a_{1}\ldots a_{n}}\left(x_{1},\ldots,x_{n}\right),
\label{eq:3-4}
\end{equation}
in which $\Omega_{\mu_{1}\ldots\mu_{n}}^{a_{1}\ldots a_{n}}\left(x_{1},\ldots,x_{n}\right)$ represents the $n$-point function of the non-Abelian gauge field minimally coupled to the scalar matter.
From a diagrammatic point of view, it includes the one-loop graphs contributing to the gauge field $n$-point functions which can be understood as
\begin{equation}
\Gamma_{\rm eff}[G]=\mathcal{S}_{\rm eff}[G]+\mathcal{S}_{\rm eff}[GG]+\mathcal{S}_{\rm eff}[GGG]+\mathcal{S}_{\rm eff}[GGGG]+\cdots.
\end{equation}
Similarly to the previous discussion for the case of fermionic field in the section \ref{sec2}, we have that the tadpole contribution vanishes $\mathcal{S}_{\rm eff}[G]=0$. Hence, in the next step, we shall start with analysis of $\mathcal{S}_{\rm eff}[GG]$ term.
The relevant Feynman rules for the interaction of the scalar matter with a non-Abelian external gauge field, depicted in Fig. 4, are readily found as below \cite{ref1,ref2}
\begin{figure}[t]
\vspace{-1.2cm}
\includegraphics[height=4.5\baselineskip]{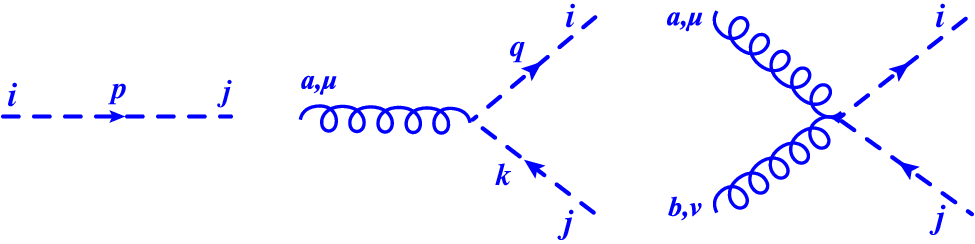}
 \centering\caption{The relevant Feynman vertices.}
\label{fig:oneloop4}
\end{figure}
\begin{equation*}
D^{ij}=\frac{i\delta^{ij}}{p^2-m^2},~~~~~\Lambda^{\mu,a}_{ij}=-ig(k+q)^{\mu}T_{ij}^{a},~~~~\Gamma^{\mu\nu,ab}_{ij}=ig^2\eta^{\mu\nu}\left(\frac{\delta^{ab}\delta^{ij}}{3}+d^{abc}T^{c}_{ij}\right).
\end{equation*}
%%%%%%%%%%%%%%%%%%%%%%%%%%%%%%%%%%%%%%
\subsection{GG-term contribution}
%%%%%%%%%%%%%%%%%%%%%%%%%%%%%%%%%%%%%%
The amplitude related with the one-loop contributions to the two-point function of the gluon can be obtained from the above Feynman rules.
There are two diagrams contributing to the GG-term at this order, which are depicted in Fig.~\ref{fig:oneloop4}.
The respective Feynman amplitudes are written as
\begin{align}
\Omega_{\mu\nu}^{ab}(p)\mid_{(a)}&=g^2\int\frac{d^{3}k}{\left(2\pi\right)^3}\frac{\left(2k-p\right)_{\mu}\left(2k-p\right)_{\nu}{\rm tr }\left(T^{a}T^{b}\right)}{\left(k^2-m^2\right)\left((k-p)^2-m^2\right)},\\
\Omega_{\mu\nu}^{ab}(p)\mid_{(b)}&=-g^2\int\frac{d^{3}k}{\left(2\pi\right)^3}\frac{\eta_{\mu\nu}\left(\delta^{ab}+d^{abc}{\rm tr }\left(T^{c}\right)\right)}{\left(k^2-m^2\right)}.
\end{align}
These expressions can be simplified by computing the traces using ${\rm tr }\left(T^{a}T^{b}\right)=\frac{1}{2}\delta^{ab}$ and ${\rm tr }(T^{a})=0$, so that we arrive at
\begin{align}
\Omega_{\mu\nu}^{ab}(p)|_{(a)}&=\frac{g^2\delta^{ab}}{2}\int\frac{d^{3}k}{\left(2\pi\right)^3}\frac{\left(2k-p\right)_{\mu}\left(2k-p\right)_{\nu}}{\left(k^2-m^2\right) \left((k-p)^2-m^2\right)},\\
\Omega_{\mu\nu}^{ab}(p)|_{(b)}&=-g^2\delta^{ab}\int\frac{d^{3}k}{\left(2\pi\right)^3}\frac{\eta_{\mu\nu}\left((k-p)^2-m^2\right)}{\left(k^2-m^2\right) \left((k-p)^2-m^2\right)}.
\end{align}
The computation of the momentum integral is straightforward using dimensional regularization. After some algebraic calculation, we consider the low-energy limit, $p^2\ll m^2$, resulting into
\begin{equation} \label{eq_GG_scalar}
\Omega_{\mu\nu}^{ab}(p)=\frac{i g^2 \delta^{ab}}{16 \pi m}\left(p_{\mu} p_{\nu}-\eta_{\mu\nu} p^2\right)+\mathcal{O}(m^{-3}),
\end{equation}
which satisfies the Ward identity $p^{\mu}\Omega_{\mu\nu}^{ab}(p)=0$.
One should observe the presence of parity even (Yang-Mills) terms only in the expression \eqref{eq_GG_scalar}.
The parity odd terms are absent in this case due to the nature of the scalar fields.
Moreover, in the next to leading order, the higher-derivative corrections to the kinetic part of YM term read as
\begin{equation} \label{eq_HD_scalar1}
\Omega_{\mu\nu}^{ab}(p)\Big|_{\tiny\mbox{HD-YM}}=\frac{i g^2 \delta^{ab}}{192 \pi m^3}\left(p_{\mu} p_{\nu}-\eta_{\mu\nu} p^2\right)p^2+\mathcal{O}(m^{-5}),
\end{equation}
actually, this contribution has the same structure as \eqref{eq_hd_YM} (obtained for fermionic matter), differing by a sign (bosonic loop instead of a fermionic loop) and also a numeric factor.
\begin{figure}[t]
\vspace{-1.2cm}
\includegraphics[height=5.5\baselineskip]{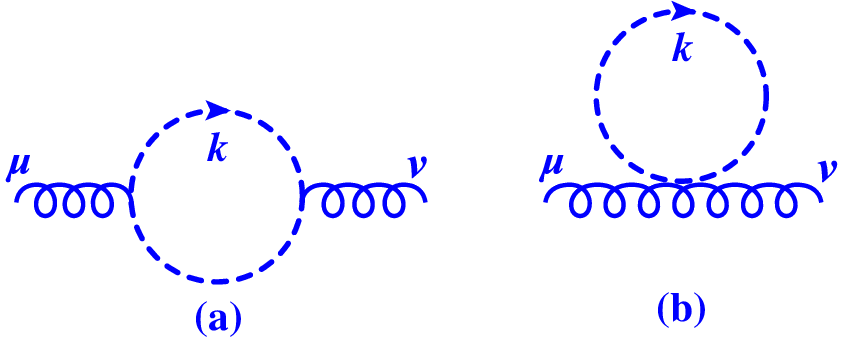}
 \centering\caption{Scalar one-loop graphs contributing to $GG$-term.}
\label{fig:oneloop4}
\end{figure}
Finally, we can cast the contribution \eqref{eq_GG_scalar} of the effective action as
\begin{equation} \label{fm3}
i \mathcal{S}_{\rm eff}[GG]=- \frac{i g^2 \delta^{ab}}{16 \pi m}\int d^{3}x \left(\partial_{\nu}G_{a}^{\mu}(x)\partial^{\nu}G_{\mu}^{b}(x)-\partial_{\nu}G_{a}^{\mu}(x)\partial_{\mu}G_{b}^{\nu}(x)\right).
\end{equation}
We observe that this expression corresponds to the kinetic term of the Yang-Mills action for the gauge field, which is corrected by the HD term \eqref{eq_HD_scalar1}.
%%%%%%%%%%%%%%%%%%%%%%%%%%%%%%%%%%%%%%%%%%%%%
\subsection{GGG-term contribution}
\begin{figure}[t]
\vspace{-1.2cm}
\includegraphics[height=7.1\baselineskip]{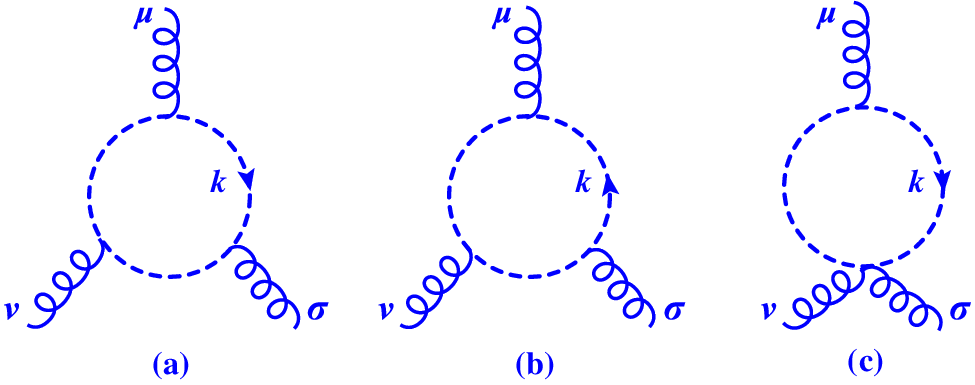}
 \centering\caption{Scalar one-loop graphs contributing to $GGG$-term.}
\label{fig:oneloop5}
\end{figure}

%%%%%%%%%%%%%%%%%%%%%%%%%%%%%%%%%%%%%%%%%%%%%
For the case of the amplitude related with the gluon three-point function there are three diagrams contributing in the scalar QCD, this is due the presence of a quartic vertex.
The corresponding graphs are depicted in Fig.~\ref{fig:oneloop5}.
Hence, using the respective Feynman rules, we can cast the diagrams (a) and (b) as
\begin{equation}
\Omega_{\mu\nu\sigma}^{abc}\Big|_{(a)+(b)}=\frac{i g^3 f^{abc}}{2}\int\frac{d^{3}k}{\left(2\pi\right)^3}\frac{\left(2k-p\right)_{\mu}\left(2k-p+r\right)_{\nu}\left(2k+r\right)_{\sigma}}
{\left(k^2-m^2\right)\left((k+r)^2-m^2\right)\left((k-p)^2-m^2\right)},
\end{equation}
while the amplitude for the diagram (c) is
\begin{equation}
\Omega_{\mu\nu\sigma}^{abc}\Big|_{(c)}=-\frac{g^3 d^{abc}}{2}\int\frac{d^{3}k}{\left(2\pi\right)^3}\frac{\eta_{\sigma\nu}\left(2k-p\right)_{\mu}}
{\left(k^2-m^2\right)\left((k-p)^2-m^2\right)}.
\end{equation}
It is easy to show that $\Omega_{\mu\nu\sigma}^{abc}\Big|_{(c)}=0$, as we would expect\footnote{As we know, the GGG-term corresponds to the 3-gluon vertex that includes only the anti-symmetric part of the structure constant $f^{abc}$.}.
Therefore, the total amplitude is simply given by
\begin{equation}
\Omega_{\mu\nu\sigma}^{abc}=\frac{i g^3 f^{abc}}{2}\int\frac{d^{3}k}{\left(2\pi\right)^3}\frac{\left(2k-p\right)_{\mu}\left(2k-p+r\right)_{\nu}\left(2k+r\right)_{\sigma}}{\left(k^2-m^2\right)\left((k+r)^2-m^2\right)\left((k-p)^2-m^2\right)}. \label{total_GGG}
\end{equation}
Gluons carry color charges and hence they are not eigenstates of the charge conjugation and one cannot expect for Furry's theorem to hold \cite{ref6,ref7,ref8}.
Therefore, the contribution of the three-point function amplitude is non-zero.

Since the amplitude \eqref{total_GGG} is finite, the loop integral is straightforward using the dimensional regularization, and in the low-energy limit $p^2,r^2,s^2\ll m^2$, it reads
\begin{equation}\label{ere}
\Omega_{\mu\nu\sigma}^{abc}= - \frac{g^3f^{abc}}{48 \pi m}\left[\eta_{\mu\sigma}\left(r-p\right)_{\nu}+\eta_{\mu\nu}\left(p-s\right)_{\sigma}+\eta_{\sigma\nu}\left(s-r\right)_{\mu}\right]+\mathcal{O}(m^{-3}).
\end{equation}
We notice that the equation \eqref{ere} corresponds exactly to the standard 3-gluon interaction term in the Yang-Mills theory.
%%%%%%%%%%%%%%%%%%%%%%%%%%%%%%%%%
Here, we notice that in the Abelian version, the total amplitude of these 3 graphs vanishes and the charge conjugation is a symmetry of the scalar QED at one-loop order. Moreover, the identity \eqref{eq:furry-abelian} is again satisfied in the scalar QED. Now, we intend to discuss Furry's theorem in the non-Abelian scalar case and understand why the total amplitude of 3-point function does not vanish.
For the scalar matter coupled to the non-Abelian gauge field, the relevant current is given by
\begin{equation}\label{eq:scalar-j-1}
{\cal{J}}_{\mu}^{a}=i\Big(\Phi^\dag T^{a}(\partial_{\mu}\Phi)-(\partial_{\mu}\Phi^\dag) T^{a}\Phi\Big).
\end{equation}
By considering the behavior of \eqref{eq:scalar-j-1} under the charge conjugation transformation and that $\Phi^{C}=\Phi^\dag$, we have
\begin{equation}\label{eq:scalar-j-2}
\big({\cal{J}}_{\mu}^{a}\big)^{C}=i\Big(\Phi^\dag (-T^{a})^{t}(\partial_{\mu}\Phi)-(\partial_{\mu}\Phi^\dag) (-T^{a})^{t}\Phi\Big).
\end{equation}
Moreover, comparing the relations \eqref{eq:scalar-j-1} and \eqref{eq:scalar-j-2}, we realize that the generators $T^a$ are replaced by $(-T^{a})^{t}$ in the conjugate representation . Hence, it is easy to see that all the aforementioned discussion in the fermionic sector works exactly here and thus for the case of anti-symmetric generators, we find
\begin{equation}
\langle \Omega\big|T\big[{\cal J}_{\mu}^{a}(x){\cal J}_{\nu}^{b}(y){\cal J}_{\rho}^{c}(z)\big]\big|\Omega\rangle \neq 0.
\label{eq:furry}
\end{equation}
While that, for the symmetric generators, the amplitude vanishes and therefore Furry's theorem is not satisfied in this case.

%%%%%%%%%%%%%%%%%%%%%%%%%%%%%%%%%
In the next to leading order, the HD contribution to the 3-gluon vertex of the YM part is readily obtained as
\begin{align} \label{eq_HD_scalar2}
\Omega_{\mu\nu\sigma}^{abc}\Big|_{\tiny\mbox{HD-YM}}=-\frac{g^3f^{abc}}{960 \pi m^3}&\bigg\{\eta_{\mu\sigma}\Big[r^{2}\left(p+r\right)-p^{2}\left(p+3r\right)+6\left(p.r\right)\left(r-p\right)\Big]_{\nu}\notag\\
&-\eta_{\nu\sigma}\Big[r^{2}\left(4r-14p\right)+p^{2}\left(p+14r\right)+\left(p.r\right)\left(6p+4r\right)\Big]_{\mu}\notag\\
&+\eta_{\mu\nu}\Big[r^{2}\left(14p+3r\right)+p^{2}\left(2p-14r\right)+\left(p.r\right)\left(4p+6r\right)\Big]_{\sigma}\bigg\}+\mathcal{O}(m^{-5}).
\end{align}
After some algebra, the cubic part \eqref{ere} of the induced effective action for the gluon is cast as
\begin{equation}\label{fm4}
i \mathcal{S}_{\rm eff}[GGG]= - \frac{i g^3 f^{abc}}{16 \pi m}\int d^{3}x~ G_{b}^{\nu}(x)G_{c}^{\mu}(x)\left(\partial_{\nu}G_{\mu,a}(x)-\partial_{\mu}G_{\nu,a}(x)\right).
\end{equation}
We observe that the gluon's amplitude \eqref{fm4} in the scalar QCD is parity even, and no Chern-Simons terms are generated (as the two-point function).
%%%%%%%%%%%%%%%%%%%%%%%%%%%%%%%%%%%%%%%%%%%%%%%%
\subsection{GGGG-term contribution}
%%%%%%%%%%%%%%%%%%%%%%%%%%%%%%%%%%%%%%%%%%%%%%%%

\begin{figure}[t]
\vspace{-1.2cm}
\includegraphics[height=6.7\baselineskip]{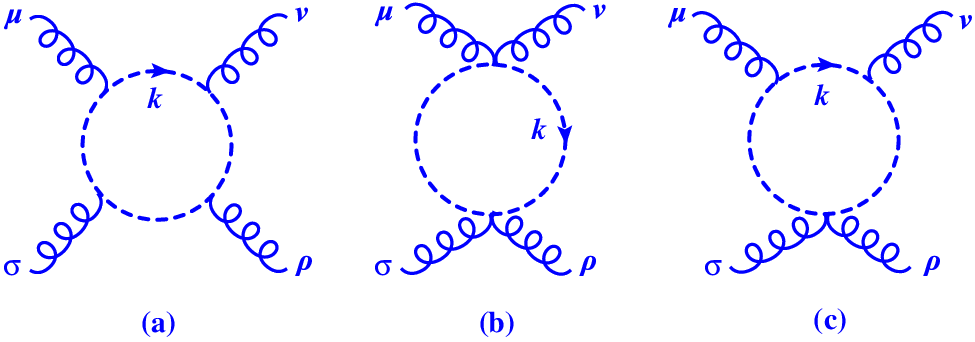}
 \centering\caption{Scalar one-loop graphs contributing to $GGGG$-term.}
\label{fig:oneloop6}
\end{figure}

According to the S-matrix expansion at the order of $g^{4}$, the four-point function contains three types of Feynman diagrams, as depicted in Fig.~\ref{fig:oneloop6}, and we have 24 different permutations (due to the external gauge field legs).
Hence, the full contribution can be formally written as
\begin{equation}\label{per}
\Omega_{\mu\nu\rho\sigma}^{abdc}=\sum_{i=1}^{24}\Big(\Omega_{\mu\nu\rho\sigma}^{abdc}\Big|_{(a,i)}+\Omega_{\mu\nu\rho\sigma}^{abdc}\Big|_{(b,i)}+
\Omega_{\mu\nu\rho\sigma}^{abdc}\Big|_{(c,i)}\Big).
\end{equation}
The amplitude of each one of the graphs (a), (b) and (c) can be computed as
\begin{subequations}
\begin{align}\label{kkk}
\Omega_{\mu\nu\rho\sigma}^{abdc}\Big|_{(a,1)}&=g^4\int\frac{d^{3}k}{\left(2\pi\right)^3}\frac{\left(2k-p\right)_{\mu}\left(2k+s\right)_{\nu}
\left(2k+2s+t\right)_{\rho}\left(2k+s+t-p\right)_{\sigma}Tr\left(T^{a}T^{b}T^{d}T^{c}\right)
}{\left[k^2-m^2\right]\left[\left(k+s\right)^2-m^2\right]\left[\left(k+s+t\right)^2-m^2\right]\left[\left(k-p\right)^2-m^2\right]},\\
\Omega_{\mu\nu\rho\sigma}^{abdc}\Big|_{(b,1)}&=g^4\int\frac{d^{3}k}{\left(2\pi\right)^3}\frac{\eta_{\mu\nu}\eta_{\rho\sigma}}
{\left[k^2-m^2\right]\left[\left(k+r+t\right)^2-m^2\right]}\left(\frac{1}{3}\delta^{ab} \delta^{dc}+\frac{1}{2}d^{abe} d^{dce}\right), \\
\Omega_{\mu\nu\rho\sigma}^{abdc}\Big|_{(c,1)}&=-g^4\int\frac{d^{3}k}{\left(2\pi\right)^3}\frac{\left(2k-p\right)_{\mu}\left(2k+s\right)_{\nu}\eta_{\rho\sigma}}{\left[\left(p-k\right)^2-m^2\right]\left[\left(k+s\right)^2-m^2\right]\left[k^2-m^2\right]}
\left(\frac{1}{6}\delta^{ab} \delta^{cd}+\frac{i}{4}f^{abe} d^{cde}+\frac{1}{4}d^{abe} d^{cde}\right),
\end{align}
\end{subequations}
where $Tr\left(T^{a}T^{b}T^{d}T^{c}\right)$ is given by \eqref{ident_T}.
Now, solving the integration by means of dimensional regularization and performing long algebraic computations in the low-energy limit $p^2,s^2,t^2,r^2\ll m^2$, we arrive at
\begin{subequations}
\begin{align}
\Omega_{\mu\nu\rho\sigma}^{abdc}\Big|_{(a,1)}&=\frac{ig^4}{96 \pi m}\left(\eta_{\mu\nu} \eta_{\rho\sigma}+\eta_{\mu\rho} \eta_{\nu\sigma}+\eta_{\mu\sigma} \eta_{\nu\rho}\right)\left(\frac{2}{3} \delta^{ab} \delta^{dc}+ d^{abe} d^{dce}+i d^{abe} f^{dce}
+if^{abe}d^{dce}-f^{abe}f^{dce}\right), \label{fff-1}\\
%%%%%%%%%%%%%%%%%%%%%%%%%%%%%%%%%%%%%%%%%%%%%%%%%%%%%%%%%%%%%%%%%%%
\Omega_{\mu\nu\rho\sigma}^{abdc}\Big|_{(b,1)}&=\frac{ig^4}{48\pi m}\eta_{\mu\nu}\eta_{\rho\sigma}\Big(2\delta^{ab} \delta^{dc}+3d^{abe} d^{dce}\Big),\label{fff-2}\\
%%%%%%%%%%%%%%%%%%%%%%%%%%%%%%%%%%%%%%%%%%%%%%%%%%%%%%%%%%%%%%%%%%%%%%%%%%
\Omega_{\mu\nu\rho\sigma}^{abdc}\Big|_{(c,1)}&=-\frac{ig^4}{32 \pi m}\eta_{\mu\nu}\eta_{\rho\sigma}\left(\frac{2}{3}\delta^{ab} \delta^{cd}+if^{abe} d^{cde}+d^{abe} d^{cde}\right).\label{fff-3}
\end{align}
\end{subequations}
Moreover, when applying 24 permutations to the results \eqref{fff-1}, \eqref{fff-2} and \eqref{fff-3}, we obtain
\begin{subequations}
\begin{align}
\Omega_{\mu\nu\rho\sigma}^{abdc}\Big|_{(a)}=& \,\frac{ig^4}{12 \pi m} \bigg[\left(\frac{2}{3}\delta^{ab} \delta^{cd}+\frac{2}{3}\delta^{ac} \delta^{db}+\frac{2}{3}\delta^{bc} \delta^{ad}+ d^{abe} d^{dce}+ d^{bde} d^{ace}+ d^{ade} d^{bce}\right)\notag\\
& \times\Big(\eta_{\mu\nu}~\eta_{\rho\sigma}+\eta_{\mu\rho}\eta_{\nu\sigma}+\eta_{\mu\sigma} \eta_{\nu\rho}\Big)\bigg], \label{qwq1}\\
%%%%%%%%%%%%%%%%%%%%%%%%%%
\Omega_{\mu\nu\rho\sigma}^{abdc}\Big|_{(b)}=& \,\frac{ig^4}{6\pi m} \bigg[\eta_{\mu\nu} \eta_{\rho\sigma}\Big(2 \delta^{ab} \delta^{cd}+3 d^{abe} d^{cde}\Big)+\eta_{\mu\rho} \eta_{\sigma\nu}\Big(2 \delta^{ad} \delta^{cb}+3d^{ade} d^{cbe}\Big)\notag\\
&+\eta_{\mu\sigma} \eta_{\nu\rho}\Big(2\delta^{ac} \delta^{bd}+3 d^{ace} d^{bde}\Big)\bigg], \label{qwq2}\\
%%%%%%%%%%%%%%%%%%%%%%%%%%%
\Omega_{\mu\nu\rho\sigma}^{abdc}\Big|_{(c)}=& \,-\frac{ig^4}{12\pi m}  \bigg[\eta_{\mu\nu} \eta_{\rho\sigma}\Big(2\delta^{ab} \delta^{cd}+3 d^{abe} d^{cde}\Big)+\eta_{\mu\rho} \eta_{\sigma\nu}\Big(2\delta^{ad} \delta^{cb}+3 d^{ade} d^{cbe}\Big)\notag\\
&+\eta_{\mu\sigma} \eta_{\nu\rho}\Big(2 \delta^{ac} \delta^{bd}+3 d^{ace} d^{bde}\Big)\bigg].\label{qwq3}
\end{align}
\end{subequations}
A complementary remark is that in the above computations some features related to the structure constants of the SU(3) group have been used \cite{ref10}, for instance
\begin{align}
&f^{ace} f^{bde}-f^{abe} f^{cde}=\frac{2}{3}\left(\delta^{ab}\delta^{cd}-\delta^{ac}\delta^{bd}\right)+d^{abe}d^{cde}-d^{ace}d^{bde}, \\
&d^{abe}d^{cde}+d^{ace}d^{bde}+d^{ade}d^{bce}=\frac{1}{3}\left(\delta^{ab}\delta^{cd}+\delta^{ac}\delta^{bd}+\delta^{ad}\delta^{bc}\right).
\end{align}
Therefore, gathering the contributions Eqs.~\eqref{qwq1}, \eqref{qwq2} and \eqref{qwq3}, the total amplitude corresponding to the 4-point function is as follows
\begin{align}\label{tot}
\Omega_{\mu\nu\rho\sigma}^{abdc}=-\frac{ig^4}{192 \pi m}\left[ f^{ace} f^{bde}\left(\eta_{\mu\nu}\eta_{\rho\sigma}-\eta_{\mu\rho}\eta_{\nu\sigma}\right)+f^{ade} f^{bce}\left(\eta_{\mu\nu}\eta_{\rho\sigma}-\eta_{\mu\sigma}\eta_{\nu\rho}\right)
+f^{abe} f^{dce}\left(\eta_{\mu\rho}\eta_{\nu\sigma}-\eta_{\mu\sigma}\eta_{\nu\rho}\right)\right].
\end{align}
As a complementary check of the above analysis, we consider the Abelian counterpart of the 4-point function
\begin{equation}
\Omega^{\mu\nu\rho\sigma}\Big|_{(a+b+c,1)}^{\rm SQED}=\frac{i e^4}{24\pi m}\left(\eta^{\mu\rho}\eta^{\nu\sigma}-2\eta^{\mu\nu}\eta^{\rho\sigma}+\eta^{\mu\sigma}\eta^{\nu\rho}\right),
\end{equation}
which vanishes when the 24 permutations are considered
\begin{equation}
\Omega^{\mu\nu\rho\sigma}\Big|_{\rm SQED}=0,
\end{equation}
as we expected.
Besides, in the next to leading order, the HD contributions to the 4-gluon YM part, arising from the general graphs, are found to be
\begin{subequations}
\begin{align}
\Omega_{\mu\nu\rho\sigma}^{abdc}\Big|_{\tiny\mbox{HD-YM}}^{(a,1)}&=\frac{ig^4}{7680 \pi m^3}\Big(\eta_{\mu\nu} \eta_{\rho\sigma}+\eta_{\mu\rho} \eta_{\nu\sigma}+\eta_{\mu\sigma} \eta_{\nu\rho}\Big)\Big(-5p^2+4s^2-5t^2+3(s.t)+35(p.s)
+30(p.t)\Big)\notag\\&\times\left(\frac{2}{3} \delta^{ab} \delta^{dc}+ d^{abe} d^{dce}+i d^{abe} f^{dce}
+i f^{abe} d^{dce}- f^{abe} f^{dce}\right)+\mathcal{O}(m^{-5}), \\
\Omega_{\mu\nu\rho\sigma}^{abdc}\Big|_{\tiny\mbox{HD-YM}}^{(b,1)}&=\frac{ig^4}{1152\pi m^3}\eta_{\mu\nu}\eta_{\rho\sigma}\Big(2\delta^{ab} \delta^{dc}+3d^{abe} d^{dce}\Big)\Big(r^{2}+t^{2}+2(r.t)\Big)+\mathcal{O}(m^{-5}), \\
\Omega_{\mu\nu\rho\sigma}^{abdc}\Big|_{\tiny\mbox{HD-YM}}^{(c,1)}&= -\frac{ig^4}{384 \pi m^3}\eta_{\mu\nu}\eta_{\rho\sigma}\left(\frac{2}{3}\delta^{ab} \delta^{cd}
+if^{abe} d^{cde}+d^{abe} d^{cde}\right)\Big(s^{2}+p^{2}-(p.s)\Big)+\mathcal{O}(m^{-5}).
\end{align}
\end{subequations}
Hence, the first amplitude of the HD contribution reads as
\begin{align} \label{eq_HD_scalar3}
&\Omega_{\mu\nu\rho\sigma}^{abdc}\Big|_{\tiny\mbox{HD-YM}}^{(a+b+c,1)}=\notag\\
&\frac{ig^4}{960 \pi m^3}\left[ \eta_{\mu\nu} \eta_{\rho\sigma}\Big\{\delta^{ab} \delta^{dc}\left(-\frac{25}{12}p^2-\frac{4}{3}s^2+\frac{5}{4}t^2+\frac{5}{3}r^2+\frac{1}{4}\left(s.t\right)+\frac{55}{12}\left(p.s\right)+\frac{5}{2}
\left(p.t\right)+\frac{10}{3}\left(r.t\right)\right)\right.\notag\\
&\left.+ d^{abe} d^{dce}\left(-\frac{25}{8}p^2-2 s^2+\frac{15}{8}t^2+\frac{5}{2}r^2+\frac{3}{8}\left(s.t\right)+\frac{55}{8}\left(p.s\right)+\frac{15}{4}
\left(p.t\right)+5\left(r.t\right)\right)\right.\notag\\
&\left.+ if^{abe} d^{dce}\left(-\frac{25}{8}p^2-2 s^2-\frac{5}{8}t^2+\frac{3}{8}(s.t)+\frac{55}{8}\left(p.s\right)+\frac{15}{4}\left(p.t\right)\right)\Big\}\right.\notag\\
&\left.+\Bigl(-5p^2+4s^2-5t^2+3\left(s.t\right)+35\left(p.s\right)+30\left(p.t\right)\Bigr)\Big\{\eta_{\mu\nu}\eta_{\rho\sigma}\left(\frac{i}{8} d^{abe} f^{dce}-\frac{1}{8} f^{abe} f^{dce}\right)\right.\notag\\
&\left.+\left(\eta_{\mu\rho} \eta_{\nu\sigma}+\eta_{\mu\sigma}\eta_{\nu\rho}\right)\left(\frac{1}{12}\delta^{ab} \delta^{dc}+\frac{1}{8} d^{abe} d^{dce}+\frac{i}{8} d^{abe} f^{dce}+\frac{i}{8} f^{abe} d^{dce}-\frac{1}{8} f^{abe} f^{dce}\right)\Big\}\right]+\mathcal{O}(m^{-5}).
\end{align}
Needless to say that the total amplitude obtained from the  24 permutations is an absurdly long expression, being omitted in the discussion.

The Eq.\eqref{tot} would generate a four gluon interaction term at order $\mathcal{O}(m^{-1})$ in the Yang-Mills effective action of the type
\begin{equation}\label{fm5}
i \mathcal{S}_{\rm eff}[GGGG]=-\frac{i g^4}{32 \pi m}\int d^{3}x~f^{abe}f^{dce}G_{\mu}^{a}(x)G_{\nu}^{b}(x)G^{d,\mu}(x)~G^{c,\nu}(x).
\end{equation}
Finally, we can collect the leading $\mathcal{O}(m^{-1})$ contributions from the one-loop order parts related to the two, three and four-point functions, Eqs. \eqref{fm3}, \eqref{fm4} and \eqref{fm5}, respectively, so that we can write the complete expression of the Yang-Mills action as
\begin{equation} \label{eff_YM2}
\Gamma_{\rm eff}^{YM}\Big|_{\mathcal{O}(m^{-1})}= -\frac{i g^2}{32 \pi m}\int d^{3}x~F_{\mu\nu}^{a}F^{a,\mu\nu},
\end{equation}
which is invariant under the Lorentz and SU(3) gauge transformations.
Naturally, the coefficients of \eqref{eff_YM2} differ from those of \eqref{eff_YM} due to the nature of the coupled matter fields.
We can observe that the HD terms for the case of scalar matter fields discussed above also generate the AAB action  \eqref{AAB}.

%%%%%%%%%%%%%%%%%%%%%%%%%%%%%%%%%%%%%%%%%%%%%%%%%%%%%%%%%%%%%%%%%%%%%%%%
\section{The gauge invariance of the higher-derivative terms}
\label{sec4}
%%%%%%%%%%%%%%%%%%%%%%%%%%%%%%%%%%%%%%%%%%%%%%%%%%%%%%%%%%%%%%%%%%%%%%%%

In order to shed some light on the general structure of the higher-derivative terms appearing in Yang-Mills-Chern-Simons action we shall apply the dimensional analysis procedure to investigate the gauge invariance of these terms.
This procedure is rather interesting because it focuses on the gauge invariant contributions.
Hence, in order to have a clear and careful discussion, we concentrate on the higher-derivative parts of the Chern-Simons and Yang-Mills terms separately below.

\begin{itemize}
  \item \emph{\textbf{Higher-derivative Chern-Simons terms}}

  %%%%%%%%%%%%%%%%%%%%%%%%%%%%%%%%%%%%%%%%%%%%%%%%%%%%%%%%%%%%%%%%%%%%%%

One should realize that since we are working in a $(2+1)$-dim. spacetime, the allowed terms in the Lagrangian density must have mass dimension 3.
Hence, one can see that the mass dimension of the gauge and fermionic fields in three dimensions is given by $[G]=\frac{1}{2}$ and $[\psi]=1$, respectively (see equation \eqref{eq:2-1}).
Thus, the mass dimension of the coupling constant would be $[g]=\frac{1}{2}$.
For the sake of simplicity in our discussion, we introduce ${\cal{G}}_{\mu}\equiv-igG_{\mu}$ that leads to $[{\cal{G}}]=1$.

The kinetic term of the ordinary CS term, arising from the leading term of the two-point function in \eqref{eff_GG1}, is proportional to $\cal{G}\partial\cal{G}$ with mass dimension 3.
We easily observe that another possible contribution (involving the gauge field) with mass dimension 3 is given by $\cal{G}\cal{G}\cal{G}$, which exactly appears as a parity-odd piece of the leading term in the analysis of the three-point function in \eqref{7}.

It is remarkable that these two types of odd-parity terms with mass dimension 3 are both at order $m^{0}$ and a special linear combination of them leads to the standard non-Abelian Chern-Simons action \eqref{eq_CS2} which is invariant under the infinitesimal gauge transformation.

In order to determine the gauge invariant higher-derivative corrections to the CS action, we must gather all  the contributions at the same order of the fermion mass $m$.
These contributions can arise from different $n$-point functions so that if we miss even one of them at any order of $m$, the gauge invariance is lost at that order.
To highlight this point, we consider the next to the leading order terms present in the two-point function in \eqref{eq_hd_CS} with odd parity, resulting in the dominant HD contribution to the kinetic term of CS at order of $m^{-2}$.
The structure of this term in the action would be as ${\cal{G}}\partial\Box{\cal{G}}$ with mass dimension of 5, compensating the coefficient $m^{-2}$, and thus resulting in a mass dimension 3 term.

Moreover, to determine the full gauge-invariant HD contributions to the CS action at order $m^{-2}$ it is necessary to consider all the possible combinations of ${\cal{G}}$ and $\partial$ with mass dimension of 5.
Due to the expansion of our results in powers of $\left(\frac{\Box}{m^{2}}\right)^{\ell}$, in general, the higher-derivative terms that may appear in $\Gamma^{(2)}$ at order of $m^{-2\ell}$ should have a structure like this ${\cal{G}}\partial\Box^{\ell}{\cal{G}}$ with mass dimension $3+2\ell$.
For example, the case of $\ell=0$ corresponds to the leading term $m^{0}$
\begin{align}
\Gamma\big|_{\tiny{m^{0}}}=a_{0}\Gamma^{(2)}_{0}+b_{0}\Gamma^{(3)}_{0},
\end{align}
which corresponds to the ordinary non-Abelian CS action with $\Gamma^{(2)}_{0}\propto{\cal{G}}\partial{\cal{G}}$ and $\Gamma^{(3)}_{0}\propto{\cal{G}}{\cal{G}}{\cal{G}}$, given in \eqref{8} and \eqref{7}, respectively. Also, the values of $a_{0}=1$ and $b_{0}=\frac{2}{3}$ are obtained via explicit analysis as given in \eqref{eq_CS2}.
Furthermore, in the next to leading order $\ell=1$, we have that
\begin{align}
\Gamma\big|_{\tiny{m^{-2}}}=a_{2}\Gamma^{(2)}_{2}+b_{2}\Gamma^{(3)}_{2}+c_2\Gamma^{(4)}_{2}+d_2\Gamma^{(5)}_{2},
\end{align}
in which
 \begin{align} \label{HD_m2}
  \Gamma^{(2)}_{2}\propto{\cal{G}}\partial\Box{\cal{G}},\quad \Gamma^{(3)}_{2}\propto{\cal{G}}{\cal{G}}\Box{\cal{G}},\quad   \Gamma^{(4)}_{2}\propto\partial{\cal{G}}{\cal{G}}{\cal{G}}{\cal{G}}, \quad
  \Gamma^{(5)}_{2}\propto{\cal{G}}{\cal{G}}{\cal{G}}{\cal{G}}{\cal{G}},
  \end{align}
  where the coefficients $a_{2},\cdots, d_2$ are determined through the explicit computation.

We should remark, however, that the terms present in \eqref{HD_m2} should be contracted with the Levi-Civit\`{a} tensor $\epsilon^{\mu\nu\rho}$ and/or the metric $\eta^{\rho\sigma}$ in order to construct Lorentz scalar objects, implying thus in many different tensorial forms contributing to the above construction.
The explicit tensorial forms of $ \Gamma^{(2)}_{2}$, $\Gamma^{(3)}_{2}$, $\Gamma^{(4)}_{2}$ are provided in our detailed results \eqref{eq_hd_CS}, \eqref{eq_hd_CS2} and \eqref{eq_hd_CS3}, respectively.

As we explained above, although we have not computed $ \Gamma^{(5)}_{2}$, it must be included to secure the gauge invariance of the full HD corrections to the CS action (in the next to leading order $m^{-2}$). Furthermore, we realize that the HD-CS action at this order includes new parity-odd self-interaction terms with four and five gluon legs.

As a conclusion, the present analysis demonstrates that in order to construct a gauge-invariant HD-CS action at order of $m^{-2\ell}$, it is necessary to consider all the relevant contributions coming from $2,3,\cdots, (3+2\ell)$-point functions.

  %%%%%%%%%%%%%%%%%%%%%%%%%%%%%%%%%%%%%%%%%%%%%%%%%%%%%%%%%%%%%%%%%%%%%
  \item \emph{\textbf{Higher-derivative Yang-Mills terms}}

Proceeding now to the analysis of the parity even contributions, we see that the YM part  are of order $m^{-1}$, while the HD corrections are of order $m^{-3}$.
Hence, we observe that YM part should receive contributions of leading order of terms with mass dimension 4, while the HD part at the next to leading order corresponds to the terms with mass dimension 6.

The kinetic term of the ordinary YM term, arising from the leading term of two-point function at  in \eqref{eff_GG1}, is proportional to $\partial\cal{G}\partial\cal{G}$ with mass dimension 4.
Other combinations with mass dimension 4 at order of $m^{-1}$ are $\partial\cal{G}\cal{G}\cal{G}$ and $\cal{G}\cal{G}\cal{G}\cal{G}$.
Indeed, these three parts are parity-even pieces of the 2, 3 and 4-point function  which sum yields the standard Yang-Mills action.

For the generic higher-derivative terms in $\Gamma^{(2)}$ at order $m^{-(2\ell+1)}$, we have that the general structure reads $\partial{\cal{G}}\partial\Box^{\ell}{\cal{G}}$ with mass dimension of $4+2\ell$. For example, the case of $\ell=0$ corresponds to the leading term $m^{-1}$
\begin{align}
\Gamma\big|_{\tiny{m^{-1}}}=a_{1}\Gamma^{(2)}_{1}+b_{1}\Gamma^{(3)}_{1}+c_{1}\Gamma^{(4)}_{1},
\end{align}
which leads to the usual YM theory with $\Gamma^{(2)}_{1}\propto\partial\cal{G}\partial\cal{G}$, $\Gamma^{(3)}_{1}\propto\partial\cal{G}\cal{G}\cal{G}$ and $\Gamma^{(4)}_{1}\propto\cal{G}\cal{G}\cal{G}\cal{G}$. Also, the coefficients $a_1, b_1$ and $c_1$ are found via our detailed computation.
In the next to leading order $\ell=1$, we obtain
 \begin{align}
\Gamma\big|_{\tiny{m^{-3}}}=a_{3}\Gamma^{(2)}_{3}+b_{3}\Gamma^{(3)}_{3}+c_3\Gamma^{(4)}_{3}+d_3\Gamma^{(5)}_{3}+e_3\Gamma^{(6)}_{3},
\end{align}
in which the invariants are
  \begin{align}
  \Gamma^{(2)}_{3}\propto\partial{\cal{G}}\partial\Box{\cal{G}},\quad \Gamma^{(3)}_{3}\propto\partial{\cal{G}}\Box{\cal{G}}{\cal{G}},\quad
  \Gamma^{(4)}_{3}\propto\Box{\cal{G}}{\cal{G}}{\cal{G}}{\cal{G}}, \cr
  \Gamma^{(5)}_{3}\propto\partial{\cal{G}}{\cal{G}}{\cal{G}}{\cal{G}}{\cal{G}}, \quad
    \Gamma^{(6)}_{3}\propto{\cal{G}}{\cal{G}}{\cal{G}}{\cal{G}}{\cal{G}}{\cal{G}}.
  \end{align}
 The tensorial form of $ \Gamma^{(2)}_{3}$, $\Gamma^{(3)}_{3}$ and $\Gamma^{(4)}_{3}$ are found in the fermionic case through \eqref{eq_hd_YM}, \eqref{eq_hd_YM2}  and \eqref{eq_hd_YM3}, as well as in the scalar  case through \eqref{eq_HD_scalar1}, \eqref{eq_HD_scalar2} and \eqref{eq_HD_scalar3}, respectively.
 Although, we have not calculated the contributions $\Gamma^{(5)}_{3}$ and $\Gamma^{(6)}_{3}$, they must be included in order to guarantee the gauge invariance of the HD corrections to YM action in the next to leading order $m^{-3}$.
 Besides, it is notable that the HD-YM action at this order includes the new parity-even self-interaction terms with five and six gluon legs (the AAB effective action \eqref{AAB}).

  Finally, in order to make a gauge-invariant HD-YM action at order of $m^{-(2\ell+1)}$, it is necessary to consider all of the relevant contributions coming from $2,3,\cdots, (4+2\ell)$-point function.

\end{itemize}

A complementary remark about the generation of the CS and YM terms can be achieved by examining the contribution arising from the momentum integrals in the low-momentum limit:

\begin{itemize}

\item In the analysis of fermionic matter, we have the trace of an odd and even number of gamma matrices appearing in the Feynman amplitudes for the $n$-point functions.

\item Moreover, the trace of an odd (even) number of gamma matrices is accompanied by an odd (even) power of fermion masses $(m, m^{3}, m^{5},...)$ [$(m^{0}, m^{2}, m^{4},...)$], which engender the Chern-Simons terms with odd parity (Yang-Mills terms with even parity).

\item In the second part, when the scalar matter is considered, there is no trace of gamma matrices, thus no parity odd terms, so that we are left only with Yang-Mills terms.

\item The presence of such mass terms can be justified as the following:
after performing the Feynman integrals in the low-energy limit, the Chern-Simons terms are produced by terms of order ${\cal{O}}(m^{-2\ell})$ while the Yang-Mills terms are produced at order ${\cal{O}}(m^{-(2\ell+1)})$.
This observation can be explicitly understood if we examine some of the standard Feynman integrals appearing in throughout of our analysis:
\begin{align}
I_4&=\int\frac{d^{3}q}{(2\pi)^{3}}\frac{q^4}{(q^{2}-\Delta)^{n}}\simeq\frac{1}{\Delta^{\left(n-\frac{3}{2}-2\right)}}, \\
I_2&=\int\frac{d^{3}q}{(2\pi)^{3}}\frac{q^2}{(q^{2}-\Delta)^{n}}\simeq\frac{1}{\Delta^{\left(n-\frac{3}{2}-1\right)}}, \\
I_0&=\int\frac{d^{3}q}{(2\pi)^{3}}\frac{1}{(q^{2}-\Delta)^{n}}\simeq\frac{1}{\Delta^{\left(n-\frac{3}{2}\right)}},
\end{align}
in which $\Delta=\Delta (p_1,p_2,...)$ is a function of the external momenta, Feynman parameters and the fermion/scalar mass.
Now, we introduce a generic form of the type $\Delta=Q^{2}+m^{2}$ to separate the contributions of the mass term from the  momenta represented generically as $Q=Q(p_1,p_2,...)$. Therefore, in the low momenta limit, we find

\begin{align}
I_4&\simeq m^{-2\left(n-\frac{3}{2}-2\right)}=m^{-2(n-3)+1}, \\
I_2&\simeq m^{-2\left(n-\frac{3}{2}-1\right)}=m^{-2(n-2)+1},\\
I_0&\simeq m^{-2\left(n-\frac{3}{2}-0\right)}=m^{-2(n-1)+1}.
\end{align}

As we see, all the low-energy integrals are proportional to an odd power of mass.
Taking this observation into account as well as the second comment, the naive and simple form of the Feynman amplitude reads
\begin{align}
 \frac{1}{m^{2s+1}}\Big[Am^{2k+1}\underbrace{Tr(\gamma^{\mu_{1}\cdots\gamma^{\mu_{2k+1}}}}_{CS-terms})
+Bm^{2k}\underbrace{Tr(\gamma^{\mu_{1}\cdots\gamma^{\mu_{2k}}})}_{YM-terms}\Big],
\end{align}
where $A=A (p_1,p_2,...)$ and $B=B (p_1,p_2,...)$ are functions of the momenta.
Finally, we can explicitly observe that the mass coefficients of the CS and YM terms are $m^{-2(s-k)}$ and $m^{-(2(s-k)+1)}$, respectively.

\end{itemize}

We summarize the conclusions of the above discussion about the perturbative generation of the CS and YM effective action in the table below.

\begin{center}
\begin{tabular}{|c|c|c|c|}
  \hline
  %after \\: \hline or \cline{col1-col2} \cline{col3-col4} ...
 \color{blue}{Order of expansion} &  \color{blue}{Induced action} & \color{blue}{spinor QCD} & \color{blue}{ scalar QCD} \\
\hline
\hspace{-0.2cm}${\cal{O}}(m^{0})$ & Chern-Simons & $\checkmark$ & $\times$ \\
    \hline
${\cal{O}}(m^{-1})$ & Yang-Mills & $\checkmark$ & $\checkmark$  \\
    \hline
${\cal{O}}(m^{-2})$ & HD Chern-Simons & $\checkmark$ & $\times$ \\
    \hline
${\cal{O}}(m^{-3})$ & HD Yang-Mills & $\checkmark$ & $\checkmark$  \\
    \hline
    \vdots & \vdots & \vdots & \vdots  \\
\hline
 ${\cal{O}}(m^{-2\ell})$ & HD Chern-Simons & $\checkmark$ & $\times$  \\
    \hline
\hspace{0.6cm}${\cal{O}}(m^{-(2\ell+1)})$ & HD Yang-Mills & $\checkmark$ & $\checkmark$  \\
    \hline
\end{tabular}
\end{center}

%%%%%%%%%%%%%%%%%%%%%%%%%%%%%%%%%%%%%%%%%%%%%%%%55555
\section{Concluding remarks}
\label{sec5}
%%%%%%%%%%%%%%%%%%%%%%%%%%%%%%%%%%%%%%%%%%%%%%%%%%%%

In this work, we have examined the perturbative generation of the effective action for the non-Abelian gauge field in a $(2+1)$-dimensional spacetime.
We have computed the two, three and four point functions in order to determine the non-Abelian Chern-Simons terms (parity odd) and Yang-Mills terms (parity even).
These terms were supplemented by the higher-derivative corrections which resulted in the Alekseev-Arbuzov-Baikov effective action (parity even) plus the HD corrections to the Chern-Simons terms (parity odd).
In order to highlight some features about the perturbative generation of the HD terms we have considered a discussion based on the dimensional analysis and gauge invariance, which allows to establish the general structure of the permissible terms.

We started by considering the perturbative generation of the non-Abelian Chern-Simons  and Yang-Mills terms when coupled to fermionic and scalar matter.
The choice of working with fermoinic and bosonic matter sources is to highlight subtle points involving the allowed terms in the effective action: in the case of fermionic matter parity odd terms are generated, while they are absent for the scalar fields.
In one hand, in the case of the Yang-Mills terms (parity even sector), they have the same structure for both fields, differing only by the numerical factors and a sign (fermionic and bosonic loop). Also, the same observation is valid for the higher-derivative terms, which correspond to the AAB effective action.
On the other hand, in the parity odd sector, only the fermionic field engender the Chern-Simons terms and their HD corrections.

In order to have a better understanding of the general structure of the effective action and the allowed terms, we followed an approach based on the dimensional analysis (mass dimension) and gauge invariance.
The starting point was to realize that the parity odd (CS) and parity even (YM) sectors are generated at order  ${\cal{O}}(m^{-2\ell})$ and  ${\cal{O}}(m^{-(2\ell+1)})$ with $\ell \geq 0$, respectively.
This observation follows from the analysis of the Feynman integrals at the low-momenta limit.
Moreover, the number of terms is very high since each allowed term should be contracted with the Levi-Civit\`{a} tensor $\epsilon^{\mu\nu\rho}$ and/or the metric $\eta^{\rho\sigma}$ in order to construct Lorentz scalar objects.
Although, one can obtain the general structure of the allowed and necessary terms at each order of the effective action, which guarantees the gauge invariance, the respective coefficients are only determined via the explicit computation.

%%%%%%%%%%%%%%%%%%%%%%%%%%%%%%%%%%%%%%%%%%%%%%%%%%%%%%%%%%%%%%%%%%%%%%%%%%%%%%%%%%%%%%%%%%%%%%%
\section*{Acknowledgments}
The authors are grateful to M.M. Sheikh-Jabbari for his valuable comments and discussions. R.B. acknowledges partial support from Conselho
Nacional de Desenvolvimento Cient\'ifico e Tecnol\'ogico (CNPq Project No. 306769/2022-0).

\end{document}